

\magnification=\magstep1
\font\sc=cmcsc10
\font\Bsc=cmcsc10 scaled \magstep 1
\font\Fnrm=cmr9
\def\fel{{\textstyle {1\over 2}}}
\def\Fel{{1\over 2}}
\def\HS{{\cal H}}
\def\Ma{{\cal M}}

\def\IN{{\cal I}}
\def\OA{{\cal A}}
\def\CB{{\cal B}}
\def\FA{{\cal F}}
\def\BO{{\cal B}({\cal H})}
\def\BOH{{\cal B}({\cal H}_\pi)}
\def\UN{{\bf 1}}
\def\NS{{\rm NS}}
\def\MT{{\tt T}}

\def\MS{{\tt S}}

\def\R{{\rm R}}

\def\Co{{\bf C}}
\def\Eg{{\bf Z}}

\def\FU{\,\vert\,}
\def\TH{{1\over 16}}

{\nopagenumbers
\line{\hfill PUPT-1406}
\line{\hfill hep-th@xxx/9306118}
\line{\hfill June 1993}
\vskip 3truecm
\centerline{\Bsc On the Quantum Symmetry of the Chiral Ising Model}

\vskip 2.5truecm
\centerline{\sc Peter Vecserny\'es\footnote{*}{\Fnrm Fulbright Fellow}}\bigskip

\centerline{\it Joseph Henry Laboratories}\smallskip

\centerline{\it Princeton University}\smallskip

\centerline{\it Jadwin Hall, Princeton, NJ 08544}

\vskip 3.5truecm
\centerline{\sc Abstract}\bigskip

\noindent We introduce the notion of rational Hopf algebras that we
think are able to describe the superselection symmetries of two dimensional
rational quantum field theories. As an example we show that a six dimensional
rational Hopf algebra $H$ can reproduce the fusion rules, the conformal
weights,
the quantum dimensions and the representation of the modular group of the
chiral Ising model. $H$ plays the role of the global symmetry algebra of the
chiral Ising model in the following sense: 1) a simple field algebra $\FA$ and
a representation $\pi$ on $\HS_\pi$ of it is given, which contains the $c=1/2$
unitary representations of the Virasoro algebra as subrepresentations; 2) the
embedding $U\colon H\to\BOH$ is such that the observable algebra $\pi(\OA)^-$
is the invariant subalgebra of $\BOH$ with respect to the left adjoint action
of $H$ and $U(H)$ is the commutant of $\pi(\OA)$; 3) there exist $H$-covariant
primary fields in $\BOH$, which obey generalized Cuntz algebra properties
and intertwine between the inequivalent sectors of the observables.

\vfill\eject}

{\bf 1. Introduction}\medskip

The Doplicher--Haag--Roberts program [1] for exploring the symmetries and
the statistics of a field theoretical model merely from `observable' data was
carried out for localized charges in Minkowski space-times of dimensions $D>2$
[2]. The set of superselection sectors $\{[\rho]\}$, which consists of the
equivalence classes $[\rho]$ of certain endomorphisms $\rho$ of the observable
algebra, can be characterized by
representations of compact groups and their statistics is restricted to Bose or
Fermi statistics. In two dimensions we expect a more rich structure in
statistics, namely, braid group statistics will play the major role. But the
possible symmetry structure, which is dual to the superselection sectors of a
model is unknown yet. Nevertheless for a distinguished class of two
dimensional field theories, for rational quantum field theories (RQFT), there
is a hope to find the corresponding symmetry algebra structure. In a RQFT by
definition [3] there are only finitely many superselection sectors and none of
them obey permutation statistics except the vacuum sector. In this case the
superselection sectors carry a unitary representation of the modular group
$\Gamma=SL(2,{\bf Z})$ even in lack of conformal symmetry and the
representations of the generators $\MS$ and $\MT$ of $\Gamma$ are given [3]
in terms of the monodromy matrix and the statistics phases of the given model,
respectively.

Since the nontrivial superselection sectors are thought to describe elementary
particles whose inner degrees of freedom are finite the corresponding symmetry
structure should have only finite dimensional irreducible representations.
Since in a RQFT there can be only a finite number of different particles that
give rise to only a finite number of superselection sectors the corresponding
symmetry algebra should be finite dimensional as well. In the final stage of
the reconstruction of a particular model we would like to have a simple field
algebra $\FA$ on which the symmetry algebra $H$ can act and the observable
algebra $\OA$ arises as the $H$-invariant subalgebra of $\FA$. The inequivalent
representations $\{\pi_r\}$ of the observables (with certain multiplicities)
arise as subrepresentations of the representation $\pi$ of $\FA$ on $\HS_\pi$
and they are in one to one correspondence with the equivalence classes of
endomorphisms $\{[\rho_r]\}$ through the vacuum representation $\pi_0$ of
$\OA$:
$\pi_r\simeq\pi_0\circ\rho_r$. Moreover, we would like to have a faithful
realization $U$ of $H$ in $\BOH$, where $U(H)$ is the commutant of $\pi(\OA)$.
But this means that $U(H)$ is a von Neumann algebra therefore $H$ should be a
finite dimensional semisimple algebra, that is a finite direct sum of full
matrix algebras. Thus the minimal central projectors of $H$ (that lead to the
inequivalent representations of $H$) are in one to one correspondence with the
inequivalent representations of the observables, because $U(H)$ and
$\pi(\OA)^-$
have a common center.

These considerations and the requirements for the representations of $H$ below
have lead us to the notion of rational Hopf algebras (RHA) that we
think are able to describe the superselection symmetries of RQFTs.\smallskip

\item{1.} Complete reducibility, finite dimensional irreducible
representations.
\item{2.} Existence of a unit representation.
\item{3.} Existence of product of representations (unique up to unitary
equivalence).
\item{4.} No loss of information taking products with the trivial
representation.
\item{5.} Notion of contragredient representation (unique up to unitary
equivalence): the product of a representation with its contragredient should
contain the trivial representation --- with multiplicity one in case of
irreducible representations.
\item{6.} Commutativity of the product of representations up to
unitary equivalence.
\item{7.} Associativity of the product of representations up to unitary
equivalence.
\item{8.} Representations of the symmetry algebra do form a braided
monoidal $C^*$-category.
\item{9.} $\hat H$, the set of equivalence classes of unitary
irreducible representations of the symmetry algebra $H$ should give rise
to a $\vert \hat H\vert$-dimensional unitary representation of the
modular group $\Gamma$.\medskip

Since we want a RHA $H$ to reproduce not only the fusion rules of the
superselection sectors $\{[\rho]\}$ of the corresponding observable algebra
$\OA$, but also their braid group representations, statistics parameters and
the monodromy matrix arising from the statistics operators [1] we are to
study algebra embeddings of $\nu\colon H\to M_n(H)$ type, that is amplifying
monomorphisms $\nu$ of $H$, where $M_n(H)$ is an $n\times n$ matrix
with entries in $H$, in order to mimic the endomorphisms $\rho\colon
\OA\to\OA$ of the observables. The composition of amplimorphisms $\{\nu\}$
leads
to an associative product $\times$. In order to describe the braiding
properties
of $\times$ one can introduce the notion of `statistics operator' in $H$, which
has properties similar to the ones of the statistics operator in algebraic
field theory [4]. Therefore the statistics operators of $H$
naturally lead to statistics parameters, coloured braid
group representations and the monodromy matrix of $H$. If we want to
associate a RHA to an observable algebra as its global symmetry algebra we
think that the analogous quantities should match.

Amplifying monomorphisms of a RHA $H$ are naturally provided by the coproduct
$\Delta\colon H\to H\otimes H$
$$H\ni a\mapsto (\UN\otimes e_r)\cdot\Delta(a)\in M_{n_r}(H),\eqno(1.1)$$
where $\UN$ is the unit element of $H$ and $e_r\in H$ is a central projector
corresponding to a simple summand of dimension $n_r$ of $H$. If an embedding of
type (1.1) is unit preserving it
gives index $n_r^2$ [5], the `minimal' embedding $H\mapsto H\otimes e^{11}$,
where $e^{11}$ is a matrix unit, gives index one. If we want to reproduce a
noninteger statistical dimension $d_r$ of a sector $[\rho_r]$, which is equal
to the square root of the corresponding index $[\OA : \rho_r(\OA)]$, we are
forced to use a unit non-preserving coproduct in $H$ that can lead to an
intermediate statistical dimension $1<d_r<n_r$. Heuristically the origin of a
possible unit non-preserving property of the coproduct can also be understood
as the contradiction of the `symmetrization' procedure one has to perform on
the tensor product basis to decompose a product of two representations of $H$
and the braiding properties of fields in the product
carrying these representation spaces. In the reconstruction of field theories
in space-time dimensions $D>2$ [2] the integer statistical dimension $d_r$
plays also the role of the dimension $n_r$ of the representation of
the symmetry group and the multiplicity of the corresponding sector of the
observables in the representation of the field algebra, that is $d_r=n_r$.
In case of two dimensional field theories the latter roles can be played
by the (integer) square root of the cardinality $n_r^2$ of a quasibasis [6]
corresponding to a conditional expectation $E_r\colon\OA\to\rho_r(\OA)$, while
the noninteger statistical dimension $d_r$ arises as the square root of the
index $d_r^2$ provided by this quasibasis. If $d_r$ is an integer we think that
these two notions should coincide, i.e. $d_r=n_r$, as in the case $D>2$.

In lack of a proof or a counterexample whether RHAs are dual to the
superselection sectors of RQFTs or not we can try to test the ability of RHAs
on particular RQFT models. Due to the lack of available examples we think that
the chiral half of unitary rational conformal models can be used since they
also have only finite number of inequivalent representations of the chiral
algebra (Virasoro, Kac--Moody, \dots, etc.) and this finite set of
representations gives rise to a finite dimensional unitary representation of
the modular group $\Gamma$.

Since chiral conformal field theories live on the non-contractible `space-time'
$S^1$ an extra difficulty arises from the point of view of algebraic field
theory: the set $\IN$ of open nondense intervals are
not directed with respect to the partial ordering given by the inclusion of
sets therefore the global observable algebra cannot be defined as the inductive
limit of the local ones. One can only speak about a consistent family of von
Neumann algebras of local observables $\{\OA(I),\, I\in\IN\}$ acting on a
Hilbert space $\HS_0$ and obeying isotony, locality and covariance with respect
to the M\"obius group [7]. In this case a representation $\pi$, by definition,
is a consistent family of representations $\{\pi_I,\, I\in\IN\}$ of the local
observable algebras on some Hilbert space $\HS_\pi$, that is $\pi_I$ is a
representation of $\OA(I)$ on $\HS_\pi$ and $J\subset I$ implies
$\pi_{J\vert \OA(I)}=\pi_I$. One can look for the universal algebra $\OA$ [7],
which plays the role of the global observable algebra in the sense that there
exists a unique representation $\pi$ of $\OA$ on $\HS_\pi$ such that
$\pi_{|\OA(I)}=\pi_I$.

Since we examine the chiral Ising model in terms of
Neveu--Schwarz (NS) and Ramond (R) Majorana fermion modes we also use this
description of the global observables: $\OA_\Ma$ is given as a direct
sum of NS and R fermion mode bilinears. But this observable algebra $\OA_\Ma$
has two (related) drawbacks: first, it is not simple, therefore its irreducible
representations are not faithful, second, it contains `nonlocal' observables
in a sense that for example the R bilinears somehow know that they can live
only
on the twofold covering of the underlying spacetime $S^1$. Of course the very
choice of this observable algebra is unavoidable if one intends to describe all
of the superselection sectors by endomorphisms of the observable algebra
$\OA_\Ma$ [8]. But we think that in a true local treatment the natural thing
would be to restrict ourselves to the net of local --- in both sense ---
observables $\{\OA(I),\, I\in\IN\}$ and to allow localized amplifying
homomorphisms $\nu\colon\OA\to M_n(\OA)$ of them.
Then it is the localized amplimorphism $\nu$ that should `cut' and `sew
together' the $n$ copies of $S^1$ on a `common' interval and `mix' their local
observables in a way that the arising irreducible representation of
$\{\OA(I),\, I\in\IN\}$ is globally meaningful only on the $n$-fold covering
$\tilde S^1$ of $S^1$. The localized transportable endomorphisms may pick up a
nontrivial monodromy by transportation with $2\pi$ thus one may obtain a
nonequivalent endomorphism to the original one. In case of localized
transportable amplimorphisms $\rho\colon\OA\to M_n(\OA)$ the equivalence can
be ensured by local observables in $M_n(\OA)$. In this
description of the R sector we would expect that $n=2$ because locally the
twofold covering of $S^1$ `looks' as doubled intervals of $S^1$.
This treatment can be supported by the description of conformal field theories
given in [9] and by the fact that even in the global description
$H$-covariance requires the use of proper amplimorphisms in case of the R
sector.

To show that the $c=1/2$ unitary Virasoro representations can be obtained
from the vacuum sector through amplifying homomorphisms of the observables we
derive $H$-covariant multiplet matrix fields, or primary fields, vertex
operators in conformal field theoretical language, that obey a slightly
modified weak F-algebra (generalized Cuntz algebra [10]) relations [11]. The
modification is due to the unit non-preserving property of the coproduct of
the global symmetry algebra $H$ of the chiral Ising model.

The organization of the paper is as follows. In Chapter 2.1 we give the
defining properties of a RHA $H$ and examine products of amplifying
monomorphisms of $H$. For the description of their braiding properties we
introduce the notion of the statistics operator in $H$ that leads to the notion
of statistics parameters and the monodromy matrix of $H$. In Chapter 2.2
RHAs are constructed that obey Ising fusion rules. We recover from $H$ the
conformal weights ({\it mod} 1), the central charge ({\it mod} 8) and the
modular group representation of the chiral Ising model. In Chapter 3 we discuss
the field theory of the chiral Ising model. We show that $H$ arises as the
commutant of the observables in the representation $\pi$ of the field
algebra $\FA$. After examining various $H$-actions on $\BOH$ $H$-covariant
primary fields are constructed, which obey generalized Cuntz algebra properties
and lead to amplifying homomorphisms of the observables. Finally, Chapter 4
contains a short discussion and an outlook.

\bigskip
{\bf 2. Rational Hopf algebras}\medskip

{\it 2.1. Statistics operators and parameters in a rational Hopf algebra}
\smallskip

Here we give a short summary about the defining properties
(points corresponding to requirements in the Introduction) and about the
construction of the statistics operator, statistics parameters and the
monodromy matrix of a rational Hopf algebra $H$. The detailed description
will be given in [12].

The defining properties are as follows:\smallskip

\item{1.} H is an associative finite dimensional semisimple ${}^*$-algebra
with unit.
\item{2.} The counit $\epsilon\colon H\to{\bf C}$ is a unit preserving
${}^*$-homomorphism.
\item{3.} The coproduct $\Delta\colon H\to H\otimes H$ is a
${}^*$-monomorphism. $(a\otimes b)^*=a^*\otimes b^*$.
\item{4.} The counit obeys the property
$$(\epsilon\otimes {id})\circ\Delta(a)=\rho a \rho^*\qquad\qquad
({id}\otimes\epsilon)\circ\Delta(a)=\lambda a\lambda^*$$
with unitaries $\rho,\lambda\in H$.
\item{5.} The antipode $S$ is a linear ${}^*$-antiautomorphism of $H$. There
exist nonzero elements $l,r\in H$ such, that for all $a\in H$
$$a^{(1)}\cdot l\cdot S(a^{(2)})=l\cdot\epsilon(a),\qquad\qquad
S(a^{(1)})\cdot r\cdot a^{(2)}
=\epsilon(a)\cdot r,$$
where $a^{(1)}\otimes a^{(2)}\equiv\Delta(a)$.
\item{6.} Quasi cocommutativity: there exists $R\in H\otimes H$ such that
$$\eqalign{\Delta'(a)\cdot R&=R\cdot\Delta(a),\qquad a\in H;\cr
\Delta'({\bf 1})\cdot R=&R=R\cdot\Delta({\bf 1}),\cr
R\cdot R^*=\Delta'({\bf 1}),\quad &~\quad R^*\cdot R=\Delta({\bf 1}),\cr}$$
where $\Delta'$ denotes the coproduct with interchanged tensor product factors.
\item{7.} Quasi coassociativity: there exists $\varphi\in H\otimes H\otimes H$
such that
$$\eqalign{(\Delta\otimes {id})\circ\Delta(a)\cdot\varphi &=
\varphi\cdot({id}\otimes\Delta)\circ\Delta(a),\qquad a\in H;\cr
(\Delta\otimes {id})\circ\Delta({\bf 1})\cdot\varphi =&\varphi=
\varphi\cdot({id}\otimes\Delta)\circ\Delta({\bf 1}),\cr
\varphi\cdot \varphi^*=(\Delta\otimes {id})\circ\Delta({\bf 1}),\quad
&~\quad \varphi^*\cdot\varphi=({id}\otimes\Delta)\circ\Delta({\bf 1}).\cr}$$
\item{8.a} Triangle identity:
$$({id}\otimes\epsilon\otimes{id})\varphi=(\lambda^*\otimes\UN)\Delta(\UN)
(\UN\otimes \rho)\, .$$
\item{8.b} Square identities:
$$S(\varphi_1)\cdot r\cdot\varphi_2\cdot l\cdot S(\varphi_3)=\UN=
\varphi_1^*\cdot l\cdot S(\varphi_2^*)\cdot r\cdot\varphi_3^*.$$
\item{8.c} Pentagon identity:
$$(\Delta\otimes {id}\otimes {id})\varphi\cdot ({id}\otimes
{id}\otimes\Delta)\varphi=(\varphi\otimes\UN)\cdot({id}\otimes\Delta\otimes
{id})\varphi\cdot(\UN\otimes \varphi)\, .$$
\item{8.d} Hexagon identities:
$$\eqalign{\varphi_{231}\cdot(\Delta\otimes {id})R\cdot\varphi_{123}
&=R_{13}\cdot
\varphi_{132}\cdot R_{23}, \cr
\varphi_{312}^*\cdot({id}\otimes\Delta)R\cdot\varphi_{123}^*&=R_{13}\cdot
\varphi_{213}^*\cdot R_{12}\, , \cr}$$
where if $\varphi\equiv\varphi_{123}=\sum\varphi_1\otimes\varphi_2
\otimes\varphi_3$ then $\varphi_{231}=\sum\varphi_2\otimes\varphi_3
\otimes\varphi_1$.\smallskip

\item{9.} The monodromy matrix $Y\in M_{\vert\hat H\vert}(H)$ of the symmetry
algebra is invertible.\medskip

We note that rational Hopf algebras share a lot of properties of
quasitriangular
quasi Hopf algebras [13] and weak quasi Hopf algebras [14] in the sense that
the coproduct is not necessarily coassociative and unit preserving,
respectively. The main difference is in the ${}^*$-algebra properties and in
the
most restrictive property 9. of rational Hopf algebras, since the latter is the
only one among the nine properties that excludes group algebras of finite
non-Abelian groups.

The representations of $H$ are $D\colon H\to M_n({\bf C})$ ${}^*$-algebra
homomorphisms. Due to 1. they are completely reducible. The matrix units
$\{e_r^{ij}\in H\, |\, r\in\hat H,\, i,j=1,\ldots ,n_r\}$
provide us a linear basis in $H$, where $\hat H$ is the index set of minimal
central projectors in $H$. The defining unitary irreducible representations
$D_r,\, r\in\hat H$ of $H$ are given as
$$D_r^{ij}(a):=a_r^{ij},\qquad
  a=\sum_{p\in \hat H}\sum_{i,j=1}^{n_p}a_p^{ij}e_p^{ij},\qquad
  a\in H,\quad a_p^{ij}\in\Co.\eqno(2.1)$$
The counit $\epsilon$ in 2. is considered as the one dimensional trivial
representation. The coproduct in 3. allows us to define product of
representations:
$$(D_1\times D_2)(a):=(D_1\otimes D_2)(\Delta(a))
\equiv D_1(a^{(1)})\otimes D_2(a^{(2)}).\eqno(2.2)$$
Since we do not require the unit preserving property for the coproduct
null-representation are allowed, but the one-dimensional null
representation is not considered to be irreducible. 4. means that product
representations with the trivial one lead to equivalent representations to
the original ones. Using the antipode in 5. one can define the contragredient
$\bar D$ of a representation $D$, namely: $\bar D(a):=D^t(Sa),
\, a\in H$, where $t$ denotes the transposition of a matrix. Properties of $S$
ensure that $\bar D\colon H\to M_n(H)$ is a ${}^*$-homomorphism, where $n$ is
the dimension of $D$. Moreover, $l$ and $r$ serve as natural
$(D\times \bar D\vert\epsilon)$ and $(\epsilon\vert \bar D\times D)$
intertwiners. Properties 6--7. ensure the commutativity and associativity
of the product of representations up to unitary
equivalence. The identities 8.a--d are responsible for the correct braided
monoidal structure of the representations. The definition of the monodromy
matrix will be given after the construction of the statistics operator.

In an other language one can say that the coproduct $\Delta$ provides us a
possibly unit non-preserving embedding of $H$ into $H\otimes H$. Of course,
we are interested in such embeddings only up to inner unitary automorphisms
of $H\otimes H$. This means that the algebra $H$ with the coproduct
$$\Delta_U(a)=U\Delta (a)U^*,\quad U\in
{\cal U}_2\equiv\{ V\in H\otimes H\, |\, VV^*=\UN\otimes\UN\}\eqno(2.3)$$
is considered to be equivalent to the original RHA if their
representations are equivalent from a category theoretical point of view. One
can show that this equivalence always holds with elements in
$U=\sum_k U_{1k}\otimes U_{2k}\equiv U_1\otimes U_2\in{\cal U}_2$ since
defining
$$\rho_U=\epsilon(U_1)U_2\rho,\qquad\lambda_U=U_1\epsilon(U_2)\lambda,
  \eqno(2.4a)$$
$$l_U=U_1lS(U_2),\qquad r_U=S(U_1^*)rU_2^*,\eqno(2.4b)$$

$$R_U=U_{21}RU_{12}^*,\qquad
  \varphi_U=U_{12}[(\Delta\otimes {id})U]
  \varphi[({id}\otimes\Delta)U^*]U_{23}^*,\eqno(2.4c)$$
$H_U\equiv(H,\epsilon,\Delta_U,\rho_U,\lambda_U,R_U,\varphi_U;S,l_U,r_U)$
also satisfies 1--9. We can use this `gauge freedom' ${\cal U}_2$
to reach a canonical form of a rational Hopf algebra. With suitable choice of
$U\in{\cal U}_2$ one achives the more familiar properties
$$(\epsilon\otimes {id})\circ\Delta={id}=({id}\otimes\epsilon)
  \circ\Delta\leqno4.'$$
$$({id}\otimes\epsilon\otimes {id})\varphi=\Delta(\UN)\leqno8.a'$$
of the counit and $\varphi$ instead of 4. and 8.a. Then one proves that
$$(\epsilon\otimes{id}\otimes {id})\varphi=\Delta(\UN)
=({id}\otimes{id}\otimes\epsilon)\varphi,\eqno(2.5a)$$
$$\epsilon(R_1)\cdot R_2=\UN=R_1\cdot\epsilon(R_2)\eqno(2.5b)$$
and $\epsilon\circ S=\epsilon$ fulfil as well. Using the remaining gauge
freedom
$l$ and $r$ can be transformed into invertible central elements of $H$ and one
of them can be even positive. Using an other gauge freedom
$$S_U(a)=US(a)U^*, \qquad l_U=lU^*,\quad r_U=Ur,\qquad
  U\in{\cal U}_1\equiv\{V\in H\,\vert\, VV^*=\UN\}\eqno(2.6)$$
one proves that $S$ can be chosen as $S(e_r^{ij})=e_{\bar r}^{ji},\, r\in\hat
H,
\, i,j=1,\ldots ,n_r$, where $r\mapsto \bar r$ is the involution describing the
`charge conjugation' among isomorphic direct summands $e_rH$ and $e_{\bar r}H$
in $H$.

To obtain information about the braiding properties of the coproduct we will
use
amplifying monomorphisms, or amplimorphisms, for short, of $H$ instead of
representations. The benefit of this choice stems from the existence of a left
inverse of an amplimorphism, which can lead to the notion of conditional
expectations, statistics parameters and index. To stress the similarity of the
following construction described in details in [12] to that of in
algebraic field theory [4] we use the same name for the corresponding
quantities.

An amplimorphism of $H$ is a ${}^*$-algebra monomorphism
$\nu\colon H\to M_n(H)$. A left inverse $\Phi_\nu\colon M_n(H)\to H$ of $\nu$
is a positive linear map having the property
$$\eqalign{\Phi_\nu(\UN_n)&=\UN,\cr
  \Phi_\nu(\nu(a)\cdot B\cdot\nu(c))&=a\cdot\Phi_\nu(B)\cdot c,\qquad a,c\in H,
  \ B\in M_n(H).\cr}\eqno(2.7)$$
The linear space of intertwiners between the amplimorphisms
$\mu\colon H\to M_m(H)$ and $\nu\colon H\to M_n(H)$ is
$$(\mu\vert\nu)=\{ T\in{\rm Mat}(m\times n, H)\vert \mu(a)T=T\nu(a),a\in H,
  \mu({\bf 1})T=T=T\nu({\bf 1})\} .$$
Amplimorphisms $\nu_1$ and $\nu_2$ are called equivalent, $\nu_1\sim\nu_2$ if
there is an equivalence $T$ in the intertwiner space $(\nu_1|\nu_2)$, that is
$$TT^*=\nu_1({\bf 1}),\qquad \quad T^*T=\nu_2({\bf 1}).$$
One can define subobjects, direct sums and product of amplimorphisms. The
latter is given by
$$(\mu\times\nu)^{i_1j_1,i_2j_2}(a):=\mu^{i_1i_2}(\nu^{j_1j_2}(a)),\qquad
  a\in H,\eqno(2.8a)$$
and it is associative. The product $T_1\times T_2$ of intertwiners
$T_i\in(\mu_i\vert\nu_i),\, i=1,2$ and is defined as
$$T_1\times T_2:=\mu_1(T_2)\cdot(T_1\otimes I_{n_2})=(T_1\otimes
  I_{m_2})\cdot\nu_1(T_2)\in(\mu_1\times\mu_2\vert\nu_n\times\nu_2).
  \eqno(2.8b)$$
An amplimorphism $\mu\colon H\to M_m(H)$ always leads to a representation: we
have only to compose it with the `vacuum representation' that is with the
trivial representation, the counit $\epsilon$:
$$D_\mu:=\epsilon\circ\mu\colon H\to M_m(\Co),\qquad
  D^{ij}_\mu(a):=\epsilon(\mu^{ij}(a)),\qquad a\in H;\ i,j=1,\ldots ,m.
  \eqno(2.9)$$
On the other hand every nonzero representation $D$ of $H$ defines a special
amplimorphism $\mu_D\colon H\to M_m(H)$ by the help of the coproduct:
$$\mu_D(a):=a^{(1)}\otimes D(a^{(2)}),\qquad a\in H,\eqno(2.10)$$
where $m$ is the dimension of the representation $D$.

We call an amplimorphism $\nu\colon H\to M_n(H)$ natural if $\nu\sim\mu_D$,
i.e. if there is a representation $D\colon H\to M_m({\bf C})$ and an
equivalence
$T\in (\mu_D\vert\nu)\subset {\rm Mat}(m\times n,H)$.
The equivalences $\mu_{D_1}\sim\nu\sim\mu_{D_2}$ imply that $D_1$ and $D_2$ are
unitary equivalent representations. A natural amplimorphism $\nu\sim\mu_D$ is
called irreducible if the representation $D$ is irreducible. The identity
amplimorphism ${id}$ is the special amplimorphism corresponding to the trivial
representation, that is to the counit:
${id}(a)\equiv\mu_\epsilon(a)=a^{(1)}\otimes \epsilon(a^{(2)})=a$.

The product of two special amplimorphisms is a natural amplimorphism since it
is given by
$$\mu_{D_1}\times\mu_{D_2}={\rm Ad}[({id}\otimes D_1\otimes D_2)\varphi]\circ
  \mu_{D_1\times D_2},\eqno(2.11a)$$
that is
$$\eqalign{\mu_{D_1}\times\mu_{D_2}(a)&=[\varphi_1\otimes D_1(\varphi_2)
  \otimes D_2(\varphi_3)]\cdot\mu_{D_1\times D_2}(a)\cdot
  [\varphi_1^*\otimes D_1(\varphi_2^*)\otimes D_2(\varphi_3^*)]\cr
  &=a^{(11)}\otimes D_1(a^{(12)})\otimes D_2(a^{(2)}).\cr}\eqno(2.11b)$$
If $\nu_1,\nu_2$ are natural amplimorphism with equivalences
$T_i\in(\nu_i|\mu_{D_i}),\, i=1,2$ then their product is natural because
$$(T_1\times T_2)\cdot \varphi_1\otimes D_1(\varphi_2)\otimes D_2(\varphi_3)\in
  (\nu_1\times\nu_2|\mu_{D_1\times D_2})\eqno(2.12)$$
is an equivalence. Therefore natural amplimorphisms are closed with respect
to the product $\times$. We stress that this product is associative by
its definition (2.8a) even if the coproduct is only quasi coassociative. In
case of special amplimorphisms the equality
$$[(\mu_1\times\mu_2)\times\mu_3](a)
  =a^{(111)}\otimes D_1(a^{(112)})\otimes D_2(a^{(12)})\otimes D_3(a^{(2)})
  =[\mu_1\times(\mu_2\times\mu_3)](a)\eqno(2.13)$$
can be seen using the pentagon identity.

The braiding of amplimorphisms is described by the statistics operator
$\varepsilon$. For special amplimorphisms $\mu_1,\, \mu_2$ corresponding to
representations $D_1$ and $D_2$ the statistics operator
$\tilde \varepsilon(\mu_1;\mu_2)$ is a unitary intertwiner
$$\eqalign{\mu_2\times\mu_1(a)\cdot\tilde \varepsilon(\mu_1;\mu_2)&=
  \tilde \varepsilon(\mu_1;\mu_2)\cdot\mu_1\times\mu_2(a),\qquad a\in H,\cr
  \tilde \varepsilon(\mu_1;\mu_2)\cdot\tilde \varepsilon(\mu_1;\mu_2)^*&=
  \mu_2\times\mu_1({\bf 1}),\cr
  \tilde \varepsilon(\mu_1;\mu_2)^*\cdot\tilde \varepsilon(\mu_1;\mu_2)&=
  \mu_1\times\mu_2({\bf 1}),\cr}\eqno(2.14)$$
which is defined as
$$\tilde \varepsilon(\mu_1;\mu_2)=[({id}\otimes D_2\otimes D_1)\varphi]\cdot
  [{id}\otimes P_{12}]\cdot[{id}\otimes (D_1\otimes D_2)(R)]\cdot
  [({id}\otimes D_1\otimes D_2)\varphi^*],\eqno(2.15)$$
where $P_{12}\colon {\bf C}^{m_1}\otimes {\bf C}^{m_2}\to
{\bf C}^{m_2}\otimes {\bf C}^{m_1}$ interchanges the tensor product factors.
In case of natural amplimorphisms $\nu_1,\, \nu_2$ the statistics operator
$\varepsilon(\nu_1,\mu_1;\nu_2,\mu_2)$ is given by
$$\varepsilon(\nu_1,\mu_1;\nu_2,\mu_2):=(T_2\times T_1)^*\cdot\tilde
  \varepsilon(\mu_1;\mu_2)\cdot (T_1\times T_2),\eqno(2.16)$$
where $T_i\in(\mu_i\vert\nu_i),\, i=1,2$ are equivalences
to the corresponding special amplimorphisms $\mu_i,\, i=1,2$.

The statistics operator obeys the properties similar to that of in algebraic
field theory:

i) $\varepsilon(\nu_1,\mu_1;\nu_2,\mu_2)$ is an equivalence from
$\nu_1\times\nu_2$ to $\nu_2\times\nu_1,$

ii) $\varepsilon(\nu_1,\mu_1;\nu_2,\mu_2)$ is independent of the choice of
the special amplimorphisms $\mu_1,\, \mu_2$, therefore we can write
$$\varepsilon(\nu_1;\nu_2):=\varepsilon(\nu_1,\mu_1;\nu_2,\mu_2),$$

iii) initial conditions
$$\varepsilon(\nu;{id})=\nu(\UN)=\varepsilon({id};\nu)\, ,$$

iv) let $\nu_i\sim\tilde \nu_i$ and $T_i\in(\tilde \nu_i\vert\nu_i)$
equivalences for $i=1,2$, then
$$\varepsilon(\tilde \nu_1;\tilde \nu_2)=(T_2\times T_1)\cdot
\varepsilon(\nu_1;\nu_2)\cdot (T_1\times T_2)^*,$$

v) for composition of natural morphisms one has the hexagonal identities
$$\eqalign{\varepsilon(\nu_1\times\nu_2;\nu_3)&=(\varepsilon(\nu_1;\nu_3)
\otimes I_2)\cdot\nu_1(\varepsilon(\nu_2;\nu_3)),\cr
\varepsilon(\nu_1;\nu_2\times\nu_3)&=\nu_2(\varepsilon(\nu_1;\nu_3))
\cdot(\varepsilon(\nu_1;\nu_2)\otimes I_3),\cr}$$

vi) $\varepsilon$ is natural, that is
$$\eqalign{\varepsilon(\nu_c;\nu_b)\cdot(T_{ca}\otimes
I_b)&=\nu_b(T_{ca})\cdot\varepsilon(\nu_a;\nu_b),\cr
\varepsilon(\nu_c;\nu_d)\cdot\nu_c(T_{db})&=(T_{db}\otimes I_c)\cdot
\varepsilon(\nu_c;\nu_b)\cr}$$
fulfils for arbitrary
$T_{ca}\in(\nu_c\vert\nu_a)$ and $T_{db}\in(\nu_d\vert\nu_b)$
intertwiners,

vii) $\varepsilon_{ab}\equiv\varepsilon(\nu_a;\nu_b)$ obeys the coloured
braid relation
$$\nu_3(\varepsilon_{12}) \cdot(\varepsilon_{13}\otimes I_2)\cdot
\nu_1(\varepsilon_{23})=(\varepsilon_{23}\otimes I_1)\cdot
\nu_2(\varepsilon_{13})\cdot(\varepsilon_{12}\otimes I_3).$$
The statistics operator $\varepsilon_\nu$ of a natural amplimorphism $\nu$ is
defined as $\varepsilon_\nu:=\varepsilon(\nu;\nu).$

The conjugate $\bar\nu$ of an amplimorphism $\nu\colon H\to M_n(H)$ is defined
as
$$\bar \nu(a):=S[\nu(S(a))]^t,\eqno(2.17)$$
where $S$ is the antipode and ${}^t$ refers to the transposed matrix.
The conjugate $\bar\mu_D$ of a special amplimorphism $\mu_D$ is natural since
one proves that $\bar\mu_D\sim\mu_{\bar D}$, where $\bar D$ is the
contragredient representation of $D$. Conjugation of amplimorphisms is
involutive up to equivalence. The conjugate $\bar T$ of an intertwiner
$T\in(\nu_1\vert\nu_2)$ is defined as $\bar T:= S[T]^t$. It is also involutive
up to equivalences. One easily proves that $\bar
T\in(\bar\nu_2\vert\bar\nu_1)$.
Thus using product of equivalences one proves that the conjugate
$\bar\nu$ of a natural amplimorphism $\nu$ is natural.

A partial isometry $P_\mu \in(\mu_{\bar D}\times\mu_D\vert {id})$ for special
amplimorphisms (with nonzero $D$) can be given as
$$P_\mu^{ij,\cdot}={1\over \sqrt{{\rm tr} D(rr^*)}}\varphi_1
  \cdot D^{ji}(\varphi_3r^*S(\varphi_2)),\qquad i,j=1,\ldots {\rm dim}\ D.
  \eqno(2.18)$$
Thus a partial isometry $P_\nu \in(\bar \nu\times\nu\vert {id})$ for a natural
amplimorphism $\nu\sim\mu_D$ can also be given. A standard left inverse
$\Phi_\nu\colon M_n(H)\to H$ of a natural amplimorphism
$\nu\colon H\to M_n(H)$ ($\nu\sim\mu_D$ for a nonzero representation $D$)
is defined as
$$\Phi_\nu(A):=P^*_\nu\cdot\bar \nu(A)\cdot P_\nu,\qquad A\in M_n(H).
  \eqno(2.19)$$
Indeed, $\Phi_\nu$ is a positive linear map having the properties (2.7).

The statistics parameter matrix $\Lambda_\nu\in M_n(H)$ and the statistical
parameter $\lambda_\nu\in H$ of a natural amplimorphism $\nu\colon H\to M_n(H)$
is
defined as
$$\Lambda_\nu:=\Phi_\nu(\varepsilon_\nu),\qquad\quad
  \lambda_\nu:=\Phi_\nu(\Lambda_\nu).\eqno(2.20)$$
One proves that the statistics parameter depends only on the equivalence class
of the corresponding amplimorphism and it is in the center of $H$. For an
irreducible amplimorphisms $\nu_r,\, r\in\hat H$ it has the form
$$\lambda_r={\omega_r\over d_r}\cdot \UN,\eqno(2.21)$$
where the pure phase $\omega_r$ is the statistics phase and the positive real
number $d_r$ is the statistical dimension of the irreducible representation
$r$.
Now we can give the definition of the monodromy matrix
$Y\in M_{\vert \hat H\vert}(H)$. It is defined as
$$Y_{rs}:=d_rd_s\cdot\Phi_r\Phi_s(\varepsilon(\nu_r;\nu_s)\cdot
  \varepsilon(\nu_s;\nu_r)), \qquad r,s\in\hat H.\eqno(2.22)$$
If has the form $Y_{rs}=y_{rs}\cdot\UN,\, y_{rs}\in\Co$ and if it is
invertible then similarly to [3] one proves that
$$V(\MS)_{rs}={1\over \vert\sigma\vert}\cdot y_{rs},\qquad
  V(\MT)_{rs}=\left({\sigma\over\vert\sigma\vert}\right)^{1\over 3}\cdot
  \delta_{rs}\omega_r,
  \qquad\quad\sigma=\sum_{r\in\hat H}d_r^2\omega_r^{-1}\eqno(2.23)$$
provides us a unitary representation $V$ of the modular group $\Gamma$.

It is easy to see that the statistics parameters and the monodromy matrix are
independent of the gauge choice in ${\cal U}_1,{\cal U}_2$, while the
statistics
operators are invariant up to unitary equivalence.

\medskip
{\it 2.2. Rational Hopf algebras with Ising fusion rules}\smallskip

In this Section we construct RHAs, whose irreducible unitary representations
obey the same fusion rules as the primary fields of the chiral Ising model.

Since the Virasoro algebra has three inequivalent unitary representations at
$c=1/2$ the RHA $H$ we are looking for should be a direct sum of three full
matrix algebras. The statistical dimensions, or quantum dimensions in conformal
field theoretical language [15], in the chiral Ising model corresponding to the
sectors of conformal weights $0,1/2$ and $1/16$ are
$d_0=1,d_1=1$ and $d_2=\sqrt{2}$,
respectively. Our ansatz is to choose the dimensions $n_r$ of the corresponding
simple direct summands $M_{n_r},\, r=0,1,2$ of $H$ as the smallest integer that
obey $d_r\leq n_r$. This choice gives $n_r=d_r$ if $d_r$ is an integer and
associates one dimensional direct summands in $H$ to Abelian sectors, or simple
currents in conformal field theoretical language. Thus
$H=M_1\oplus M_1\oplus M_2$ as a ${}^*$-algebra\footnote{*}{\Fnrm After this
work was completed K. Szlach\'anyi called into our attention the preprint of
V. Schomerus [16] where in terms of Clebsch--Gordan coefficients a six
dimensional weak quasitriangular quasi Hopf algebra is given as a possible
example that describes the fusion rules of the chiral Ising model.} and
its unit is the sum of the primitive central idempotents: $\UN=e_0+e_1+e_2.$
The central projector $e_0$ of the first $M_1$ is chosen to be an integral in
$H$, that is $ae_0=\epsilon (a)\cdot e_0,\, a\in H$ [17], thus using the
matrix units as a linear basis of $H$ the counit is given as
$$\epsilon(f)=\left\{\eqalign{&1,\ f=e_0;\cr &0,\ f=e_1,e_2^{11},e_2^{12},
e_2^{21},e_2^{22}.\cr}\right.\eqno(2.24)$$
In order to reproduce the fusion rules of the chiral Ising model the nonzero
fusion coefficients in the decomposition of products of irreducible
representations of $H$ should read as
$$N_{00}^0=N_{11}^0=N_{22}^0=N_{01}^1=N_{10}^1=N_{22}^1=N_{02}^2=N_{20}^2=
N_{12}^2=N_{21}^2=1\, ,\eqno(2.25)$$
where $0,1,2$ refer to the primitive central idempotents $e_0,e_1,e_2$.
The coproduct gives an embedding of $H$ into the semisimple matrix algebra
$$H\otimes H=M_1^{(00)}\oplus M_1^{(01)}\oplus M_1^{(10)}\oplus M_1^{(11)}
\oplus M_2^{(02)}\oplus M_2^{(20)}\oplus M_2^{(12)}\oplus M_2^{(21)}\oplus
M_4^{(22)}$$
and the fusion coefficient $N_{pr}^s$ is just the number of lines in the
corresponding Bratteli diagram between the simple summands $e_sH\mapsto
e_pH\otimes e_rH$. The most general coproduct, which is consistent with the
fusion rules (2.25) and with axioms 3. and 4'., is given on the basis
elements as
$$\eqalign{\Delta(e_0)&=e_0\otimes e_0+e_1\otimes e_1+
U_{22}(e_2^{11}\otimes e_2^{11})U_{22}^*,\cr
\Delta(e_1)&=e_0\otimes e_1+e_1\otimes e_0+
U_{22}(e_2^{22}\otimes e_2^{22})U_{22}^*,\cr
\Delta(e_2^{ij})&=e_0\otimes e_2^{ij}+e_2^{ij}\otimes e_0+
U_{12}(e_1\otimes e_2^{ij})U_{12}^*+U_{21}(e_2^{ij}\otimes e_1)U_{21}^*,
\quad i,j=1,2;\cr}\eqno(2.26)$$
with unitaries $U_{12},U_{21}$ and $U_{22}$ in the simple direct summands
$M_2^{(12)}, M_2^{(21)}$ and $M_4^{(22)}$, respectively. Using the gauge
freedom ${\cal U}_2$ we can transform the $U$-s into unit matrices reaching a
cocommutative coproduct. We note that this coproduct is nothing else than the
coproduct on the group algebra ${\bf C}S_3$ truncated by the projection
$$\Delta(\UN)=\UN_{00}+\UN_{01}+\UN_{10}+\UN_{11}+
\UN_{02}+ \UN_{20}+\UN_{12}+\UN_{21}+
e_2^{11}\otimes e_2^{11}+e_2^{22}\otimes e_2^{22},\eqno(2.27)$$
where we used the notation $\UN_{rs}=e_r\otimes e_s$. The reached coproduct is
not coassociative in the direct summands $M_4^{(122)}, M_4^{(221)}$ and
$M_8^{(222)}$. Therefore we have to introduce a nontrivial associator
$\varphi$.
The most general $\varphi$ that satisfies axioms 7., 8.a' and 8.c can be
written in the form
$$\eqalignno{\varphi
&=\UN_{000}+\UN_{001}+\UN_{002}+\UN_{010}+\UN_{011}+\UN_{012}+\UN_{020}
 +\UN_{021}+\UN_{100}+\UN_{101}+\UN_{102}\cr
&+\UN_{110}+\UN_{111}+\UN_{120}+\UN_{200}+\UN_{201}+\UN_{210}
 +\omega_1\cdot \UN_{112}-\UN_{121}+\omega_1^*\cdot \UN_{211}\cr
&+e_0\otimes e_2^{11}\otimes e_2^{11}+e_0\otimes e_2^{22}\otimes e_2^{22}
 +e_2^{11}\otimes e_2^{11}\otimes e_0+e_2^{22}\otimes e_2^{22}\otimes e_0\cr
&+e_2^{11}\otimes (e_0-e_1)\otimes e_2^{11}
 +e_2^{22}\otimes (e_0+e_1)\otimes e_2^{22}\cr
&+\omega_2\cdot e_1\otimes e_2^{12}\otimes e_2^{12}
 +\omega_1^*\omega_2^*\cdot e_1\otimes e_2^{21}\otimes e_2^{21}&(2.28)\cr
&-\omega_1\omega_2\cdot e_2^{12}\otimes e_2^{12}\otimes e_1
 -\omega_2^*\cdot e_2^{21}\otimes e_2^{21}\otimes e_1\cr
&+{\alpha\over\sqrt{2}}\cdot[
  e_2^{11}\otimes e_2^{11}\otimes e_2^{11}
 +e_2^{12}\otimes e_2^{11}\otimes e_2^{21}
 -\omega_1\omega_2\cdot (e_2^{11}\otimes e_2^{12}\otimes e_2^{12}
                        +e_2^{12}\otimes e_2^{12}\otimes e_2^{22})\cr
&~\qquad +\omega_1^*\omega_2^*\cdot (e_2^{21}\otimes e_2^{21}\otimes e_2^{11}
                        +e_2^{22}\otimes e_2^{21}\otimes e_2^{21})
         +e_2^{21}\otimes e_2^{22}\otimes e_2^{12}
         +e_2^{22}\otimes e_2^{22}\otimes e_2^{22}],\cr}$$
where the parameters $\omega_1,\omega_2$ are pure phases and $\alpha=\pm 1$.
The most general $R$-matrix that satisfies axiom 6., 8.d and the equalities
(2.5b) is
$$\eqalign{R&=e_0\otimes e_0+e_0\otimes e_1+e_1\otimes e_0-
e_1\otimes e_1+e_0\otimes e_2+e_2\otimes e_0\cr
&+i\alpha_1( e_1\otimes e_2+e_2\otimes e_1)+
\omega_1^{(22)}\cdot e_1^{11}\otimes e_2^{11}+
\omega_2^{(22)}\cdot e_2^{22}\otimes e_2^{22},\cr}\eqno(2.29a)$$
where
$$\omega_1^{(22)}=e^{-i{\pi\over 8}
[\alpha_1-2(1-\alpha)-4(1-\alpha_2)]},\qquad
\omega_2^{(22)}=e^{i{\pi\over 8}
[3\alpha_1+2(1-\alpha)+4(1-\alpha_2)]},\eqno(2.29b)$$
and $\alpha_1,\alpha_2$ can have the values $\pm 1$.

Since the sectors of the Ising model are selfconjugate on the basis of the
gauge
freedom ${\cal U}_1$ in (2.6) we choose the transposition as the antipode. Then
the general solution for the intertwiners $l,r\in H$ that obey 5.
is given by
$$l=l_0\cdot e_0+l_1\cdot e_1+l_2\cdot e_2^{11},\qquad
  r=r_0\cdot e_0+r_1\cdot e_1+r_2\cdot e_2^{11},\qquad
  l_p,r_p\in\Co,\ p=0,1,2.\eqno(2.30)$$
The square identities imply the relations
$$r_0l_0=1,\qquad r_1l_1=1,\qquad r_2l_2=\alpha \sqrt{2}.\eqno(2.31)$$
We make the choice
$$l_0=r_0=l_1=r_1=1,\qquad\quad \alpha l_2=r_2=2^{1\over 4}.\eqno(2.32)$$
Using the remaining gauge freedom in ${\cal U}_2$ that leaves the coproduct and
the intertwiners $\lambda,\rho,l,r,R$ fix we can transform the phases
$\omega_1,\omega_2$ in (2.28) into one. Thus we have found eight inequivalent
$H=M_1\oplus M_1\oplus M_2$ algebras obeying Ising fusion rules and satisfying
axioms 1--8. Using a ${\cal U}_2$ gauge transformation, which is nontrivial
only in the $e_2H\otimes e_2H$ summand
$$U_{22}=\left(\matrix{{1\over \sqrt{2}}&0&0&{-1\over \sqrt{2}}\cr
0&1&0&0\cr 0&0&1&0\cr
{1\over \sqrt{2}}&0&0&{1\over \sqrt{2}}\cr}\right)\in M_4^{(22)},\eqno(2.33)$$
and applying the  transformation properties (2.4) we reach the canonical form
of these algebras, where $l$ and $r$ are invertible central elements:

$$\epsilon(f)=\left\{\eqalign{&1,\ f=e_0;\cr &0,\ f=e_1,e_2^{11},e_2^{12},
  e_2^{21},e_2^{22};\cr}\right.\eqno(2.34a)$$
$$f^*=\left\{\eqalign{&f,\ f=e_0,e_1,e_2^{11},e_2^{22};\cr
                      &e_2^{21},\ f=e_2^{12};\cr
                      &e_2^{12},\ f=e_2^{21};\cr}\right.\qquad\quad
  S(f)=\left\{\eqalign{&f,\ f=e_0,e_1,e_2^{11},e_2^{22};\cr
                     &e_2^{21},\ f=e_2^{12};\cr
                     &e_2^{12},\ f=e_2^{21};\cr}\right.\eqno(2.34b)$$

$$\eqalign{\Delta(e_0)&=e_0\otimes e_0+e_1\otimes e_1+\fel\cdot
  [e_2^{11}\otimes e_2^{11}+e_2^{12}\otimes e_2^{12}+e_2^{21}\otimes e_2^{21}
  +e_2^{22}\otimes e_2^{22}],\cr
  \Delta(e_1)&=e_0\otimes e_1+e_1\otimes e_0+\fel\cdot
  [e_2^{11}\otimes e_2^{11}-e_2^{12}\otimes e_2^{12}-e_2^{21}\otimes e_2^{21}
  +e_2^{22}\otimes e_2^{22}],\cr
  \Delta(e_2^{ij})&=e_0\otimes e_2^{ij}+e_2^{ij}\otimes e_0
                 +e_1\otimes e_2^{ij}+e_2^{ij}\otimes e_1;\qquad i,j=1,2;\cr}
  \eqno(2.34c)$$

$$\eqalign{
  R&=e_0\otimes e_0+e_0\otimes e_1+e_1\otimes e_0-e_1\otimes e_1\cr
   &+e_0\otimes e_2+e_2\otimes e_0
    +i\alpha_1\cdot (e_1\otimes e_2+e_2\otimes e_1)\cr
   &+{\omega\over\sqrt{2}}\cdot
     [e_2^{11}\otimes e_2^{11}+e_2^{22}\otimes e_2^{22}
    -i\alpha_1\cdot (e_2^{12}\otimes e_2^{12}+e_2^{21}\otimes e_2^{21})];\cr}
  \eqno(2.34d)$$

$$\eqalign{\varphi
  &=(e_0+e_1)\otimes (e_0+e_1)\otimes (e_0+e_1)\cr
  &+(e_0+e_1)\otimes (e_0+e_1)\otimes e_2
 +e_2\otimes (e_0+e_1)\otimes (e_0+e_1)\cr
&+e_0\otimes e_2\otimes e_0+e_0\otimes e_2\otimes e_1+
  e_1\otimes e_2\otimes e_0-e_1\otimes e_2\otimes e_1\cr
&+e_0\otimes e_2^{11}\otimes e_2^{11}
 +e_0\otimes e_2^{22}\otimes e_2^{22}
 +e_2^{11}\otimes e_2^{11}\otimes e_0
 +e_2^{22}\otimes e_2^{22}\otimes e_0\cr
&-e_1\otimes e_2^{11}\otimes e_2^{11}
 +e_1\otimes e_2^{22}\otimes e_2^{22}
 +e_2^{11}\otimes e_2^{11}\otimes e_1
 -e_2^{22}\otimes e_2^{22}\otimes e_1\cr
&+e_2^{11}\otimes e_0\otimes e_2^{11}
 +e_2^{22}\otimes e_0\otimes e_2^{22}
 -e_2^{12}\otimes e_1\otimes e_2^{12}
 -e_2^{21}\otimes e_1\otimes e_2^{21}\cr
&+{\alpha\over\sqrt{2}}\cdot[
  e_2^{11}\otimes e_2^{11}\otimes e_2^{11}
 -e_2^{11}\otimes e_2^{12}\otimes e_2^{12}
 +e_2^{21}\otimes e_2^{21}\otimes e_2^{11}
 +e_2^{12}\otimes e_2^{11}\otimes e_2^{21}\cr
&{}\qquad\ +e_2^{22}\otimes e_2^{22}\otimes e_2^{22}
 -e_2^{12}\otimes e_2^{12}\otimes e_2^{22}
 +e_2^{22}\otimes e_2^{21}\otimes e_2^{21}
 +e_2^{21}\otimes e_2^{22}\otimes e_2^{12}];\cr}\eqno(2.34e)$$

$$r=e_0+e_1+2^{-{1\over 4}}\cdot e_2,\qquad\quad
  l=e_0+e_1+2^{-{1\over 4}}\alpha\cdot e_2,\eqno(2.34f)$$
where $\alpha,\alpha_1,\alpha_2=\pm 1$ and
$\omega=\exp[(i\pi/8)(\alpha_1+2(1-\alpha)+4(1-\alpha_2))]$.

Now we turn to the computation of the statistics parameters and the monodromy
matrix. Using the defining irreducible representations for the special
amplimorphisms $\mu_r,\, r=0,1,2$ the corresponding statistics operators are
given as:
$$\tilde\varepsilon_{00}=
  \tilde\varepsilon_{01}=
  \tilde\varepsilon_{10}=\UN
=-\tilde\varepsilon_{11},$$

$$\tilde\varepsilon_{02}=
  \tilde\varepsilon_{20}=\left(\matrix{
 e_0+e_1+e_2^{11}&0\cr
 0&e_0+e_1+e_2^{22}\cr}\right),$$

$$\tilde\varepsilon_{12}=i\alpha_1\left(\matrix{
 e_0-e_1&-e_2^{12}\cr
 e_2^{21}&e_0-e_1\cr}\right),\quad
\tilde\varepsilon_{21}=i\alpha_1\left(\matrix{
 e_0-e_1&e_2^{12}\cr
 -e_2^{21}&e_0-e_1\cr}\right),$$

$$\tilde\varepsilon_{22}={\omega\over\sqrt{2}}\left(\matrix{
 e_0+e_1+(1+i\alpha_1)e_2^{11}&0&0
&-i\alpha_1(e_0-e_1)\cr
0&(1+i\alpha_1)e_2^{11}&0&0\cr
0&0&(1-i\alpha_1)e_2^{22}&0\cr
 -i\alpha_1(e_0-e_1)&0&0
&e_0+e_1+(1-i\alpha_1)e_2^{22}\cr}\right)$$
Since the defining irreducible representations are exactly self-conjugate,
$\bar D_r\equiv D_r,\, r=0,1,2;$ one can easily compute the corresponding
left inverses:
$$\Phi_0={id}_H;\qquad
\Phi_1(f)=\left\{\eqalign{&e_1,\ f=e_0;\cr
                          &e_0,\ f=e_1;\cr
                          &f,\ f=e_2^{ij};\cr}\right.\qquad
\Phi_2([f])=\left\{\eqalign{
&\fel\cdot e_2^{ij},\ [f]=[e_0]^{i,j};\cr
&\fel\cdot e_2^{ij},\ [f]=[e_1]^{i,j};\cr
&\fel\cdot (e_0+e_1),\ [f]=[e_2^{11}]^{1,1};\cr
&\fel\cdot (e_0-e_1),\ [f]=[e_2^{12}]^{1,2};\cr
&\fel\cdot (e_0-e_1),\ [f]=[e_2^{21}]^{2,1};\cr
&\fel\cdot (e_0+e_1),\ [f]=[e_2^{22}]^{2,2};\cr
&0,\ {\rm otherwise,}\cr}\right.$$
with $i,j=1,2$. Therefore the statistics parameter matrices read as
$$\Lambda_0=\UN,\qquad \Lambda_1=-\UN,\qquad
\Lambda_2={\omega\over\sqrt{2}}\cdot
\left(\matrix{e_0+e_1+e_2^{11}&0\cr
              0&e_0+e_1+e_2^{22}\cr}\right).$$
Applying the left inverses again one obtains the statistics parameters
$\lambda_r$, dimensions $d_r$ and phases $\omega_r$ of the irreducible
amplimorphisms $\mu_r,\, r=0,1,2$:
$$\matrix{
&\lambda_0=\UN,\qquad&\lambda_1=-\UN,\qquad
&\lambda_2={\omega\over\sqrt{2}}\cdot\UN,\cr\cr
&d_0=1,\qquad      &d_1=1,\qquad      &d_2=\sqrt{2},\cr\cr
&\omega_0=1,\qquad &\omega_1=-1,\qquad &\omega_2=\omega.\cr}$$
{}From the statistics operators $\tilde \varepsilon(\mu_r;\mu_s)$ one computes
the monodromy matrix $Y\in M_{|\hat H|}(H)$:
$$Y=\UN\otimes\left(\matrix{
1&1&\sqrt{2}\cr
1&1&-\sqrt{2}\cr
\sqrt{2}&-\sqrt{2}&0\cr}\right).$$
It is invertible, thus (2.23) leads to a unitary representation of the modular
group $\Gamma$ in $M_3(\Co)$ and we have obtained eight inequivalent rational
Hopf algebras corresponding to the eight possible choice of
$\alpha,\alpha_1,\alpha_2=\pm 1$.
All of these algebras have the same fusion rules, the same statistical
dimensions and monodromy matrix but the statistics phase
$\omega_2=\exp[(i\pi/8)(\alpha_1+2(1-\alpha)+4(1-\alpha_2))]$ is
different, namely, it can be any of the primitive 16th root of unity, that is
$\omega_2=\exp [2\pi i(2n+1)/16],\, n=0,\ldots, 7$. Analyzing the statistics
weights $w_r:=(1/2\pi i)\cdot\log \omega_r=(2n+1)/16\in [0,1)$ and the `central
charge' $c:=(-8/2\pi i)\cdot\log(\sigma/\vert\sigma\vert)=(2n+1)/2\in[0,8)$ of
these RHAs
$$\matrix{
{}&\quad w_0\quad&\quad w_1\quad&\quad w_2\quad&\quad c\quad\cr\cr
H({\rm Ising}):\qquad&\quad 0\quad&\quad 1/2\quad&\quad 1/16\quad&\quad 1/2\cr
H(\hat A_1(2)):\qquad&\quad 0\quad&\quad 1/2\quad&\quad 3/16\quad&\quad 3/2\cr
H(\hat E_8(2)):\qquad&\quad 0\quad&\quad 1/2\quad&\quad15/16\quad&\quad 15/2\cr
H(\hat B_r(1)):\qquad&\quad 0\quad&\quad 1/2\quad&\quad (2r+1)/16\quad
           &\quad (2r+1)/2\cr}$$
we can `identify' the different rational Hopf algebras as the symmetry algebras
of the chiral Ising model, the level two $SU(2), E_8$ and level one $SO(2r+1)$
Kac--Moody algebras or chiral WZW theories. All of these chiral field theories
contain three ineqivalent irreducible unitary representations of the
corresponding chiral observable algebra. They obey the same fusion rules,
the same statistics dimensions as the chiral Ising model, the representation of
the
modular generator $\MS$ is the same, only the conformal weights and the
Virasoro central charges are different and they correspond to the values in
the table, $mod\, 1$ and $mod\, 8$, respectively.

\bigskip
{\bf 3. The field algebra of the chiral Ising model}\medskip

{\it 3.1. Realization of the symmetry algebra $H$}\smallskip

The field theory of the chiral Ising model is described in [8] by real NS and
R Majorana fields on the twofold covering $\tilde S^1$ of the compactified
light cone $S^1$. The NS and R fields are `periodic' and `antiperiodic' on
$S^1$, respectively. In terms of fermion modes this universal Majorana
algebra $\Ma$ is the unital ${}^*$-algebra given by the generators
$$\UN,\,Y,\,B_n,\, n\in\fel\Eg,\qquad Y^*=Y,\ B_n^*=B_{-n}\eqno(3.1a)$$
together with the relations
$$\{ B_n,B_m\}=\fel\delta_{n+m,0}[\UN+(-1)^{2n}Y],\qquad
B_nY=(-1)^{2n}B_n=YB_n,
  \qquad Y^2=\UN.\eqno(3.1b)$$
$\Ma$ is a direct sum of the simple ${}^*$-algebras NS and R, the corresponding
projections are $(\UN\mp Y)/2$, respectively. Choosing a polarization, that is
a prescription of creation and annihilation operators, the NS and R algebras
have only one faithful unitary irreducible representations up to unitary
equivalence [18]. These are the Fock representations $\pi_\NS$ and $\pi_\R$ on
the Hilbert spaces $\HS_\NS$ and $\HS_\R$ characterized by the cyclic vacuum
vectors
$\Phi_\NS\in\HS_\NS$ and $\Phi_\R\in\HS_\R$:
$$\eqalign{
  \pi_\NS(B_n)\Phi_\NS\equiv b_n\Phi_\NS&=0,\quad n>0,\ n\in\Eg+\fel,
                      \qquad \pi_\NS(Y)\Phi_\NS=-\Phi_\NS,\cr
  \pi_\R(B_n)\Phi_\R\equiv b_n\Phi_\R&=0,\quad n>0,\ n\in\Eg,
                \qquad\qquad \pi_\R(Y)\Phi_\R=\Phi_\R,\cr}
  \eqno(3.2)$$
The observable algebra $\OA_\Ma$ is a direct sum generated by of NS and R
bilinears. The Hilbert space $\HS=\HS_\NS\oplus\HS_\R$ is decomposed into four
irreducible representations of the observable algebra $\OA_\Ma$
$$\HS=(\HS_0\oplus\HS_{1\over 2})\oplus(\HS_\TH\oplus\HS_\TH)\eqno(3.3)$$
characterized by the subscripts which are the conformal weights, because one
can
build the $c=1/2$ unitary representations of the Virasoro algebra in terms of
infinite sums of normal ordered fermion bilinears. The irreducible subspaces in
(3.3) carry the three inequivalent representations $\pi_0,\pi_{1\over 2}$ and
$\pi_\TH$ of the observable algebra.

In order to show that the rational Hopf algebra $H$ constructed in the previous
Chapter is the global symmetry algebra of the chiral Ising model
let us first describe the field theory on the Hilbert space $\HS$ in (3.3).
For an easy treatment we give an explicit equivalent realization of $\BO$ by
embedding $\Ma$ into the simple ${}^*$-algebra $\FA=M_2(\NS)$, which consists
of two by two matrices with entries in the NS algebra.

The ${}^*$-monomorphisms $\iota\colon\Ma\to M_2(\NS)$ is given as
$$\iota(B_n)=\left\{\eqalign{
&\left(\matrix{B_nB_+B_-&0\cr
              0&B_nB_-B_+\cr}\right),\ n\in\pm({\bf N}+\fel);\cr
&\left(\matrix{0&B_+\cr
              0&0\cr}\right),\ n=\fel;\cr
&\left(\matrix{0&0\cr
              B_-&0\cr}\right),\ n=-\fel;\cr
&\left(\matrix{B_{n\pm\Fel}B_-B_+&0\cr
              0&B_{n\pm\Fel}B_+B_-\cr}\right),\ n\in\pm{\bf N};\cr
&{1\over\sqrt{2}}\cdot\left(\matrix{0&B_-\cr
                  B_+&0\cr}\right),\ n=0;\cr}\right.\eqno(3.4)$$
where $B_\pm=B_{\pm\Fel}$. Thus the universal Majorana algebra is embedded
into $\FA=M_2(\NS)$, since we have constructed a direct sum image of R and NS
in $\FA$. The orthogonal projections $Y_\mp=(\UN_2\mp \iota(Y))/2$ to the
images of the NS and R algebras are given as
$$Y_-= \left(\matrix{B_+B_-&0\cr
                  0&B_-B_+\cr}\right),\qquad
  Y_+= \left(\matrix{B_-B_+&0\cr
                  0&B_+B_-\cr}\right),\eqno(3.5)$$
where $\UN_2=\UN_\NS\otimes I_2$ is the unit element of $M_2(\NS)$.

The faithful unitary irreducible representation $\pi$ of $\FA$ is given by
$\pi_\NS\otimes id_2$ on $\HS_\pi=\HS_\NS\otimes \Co^2$, where $\HS_\NS$ is
the Fock representations of the NS algebra. Since the commutants
$\iota(\NS)\cap\iota(\NS)'$ and $\iota(\R)\cap\iota(\R)'$ are both trivial the
representations $\pi_{|\NS}:=\pi\circ\iota_{|\NS}$ on the Hilbert
space $\pi(Y_-)\HS_\pi$ and $\pi_{|\R}:=\pi\circ\iota_{|\R}$ on
$\pi(Y_+)\HS_\pi$ are both irreducible. Thus $\pi_{|\NS}$ and $\pi_{|\R}$ are
unitary equivalent to the Fock representations  $\pi_\NS$ and $\pi_\R$ in
(3.2), respectively. Restricting ourselves to the observable subalgebra
$\OA\equiv\iota(\OA_\Ma)$ the total Hilbert space $\HS_\pi$ is decomposed into
four irreducible representations as before
$$\HS_\pi=(\HS_0\oplus\HS_\Fel)\oplus (\HS^1_\TH\oplus\HS^2_\TH).\eqno(3.6)$$
The corresponding cyclic vacuum vectors with respect to the observables are
given as
$$\eqalign{
 &\Omega_0=\Omega_\NS=\vert 0\rangle\otimes f_1\in\HS_0,\qquad
  \Omega_1=\pi_{|\NS}(B_-)\Omega_\NS=\vert\fel\rangle\otimes f_2\in\HS_\Fel,\cr
 &\Omega_2^1=\Omega_\R=\vert 0\rangle\otimes f_2\in\HS^1_\TH,\qquad
  \Omega_2^2=\pi_{|\R}(\sqrt{2}B_0)\Omega_\R=\vert\fel\rangle\otimes
   f_1\in\HS^2_\TH,\cr}\eqno(3.7)$$
where $\vert 0\rangle$ is the Fock vacuum in $\HS_\NS$,
$\vert\Fel\rangle=b_-\vert 0\rangle$ and $\{ f_1,f_2\}$ is an orthonormal basis
in ${\bf C}^2$.

Now the commutant of $\pi(\OA)$ in $\BOH$, which is isomorphic
to the semisimple matrix algebra $M_1\oplus M_1\oplus M_2$, can be given
explicitely. If $P_\pm\colon\HS_\NS\to\HS_\NS$ denote the
projections $(\UN_\NS\pm (-1)^{\bf F})/2\in\CB(\HS_\NS)$, where ${\bf F}$
is the fermion number operator, then a unit preserving ${}^*$-monomorphism
$U\colon H\to\BOH$ can be given by defining the the images of matrix units
of the rational Hopf algebra $H$ in (2.34) as
$$\eqalign{
U(e_0)&=P_+b_+b_-\otimes f^{11}+P_+b_-b_+\otimes f^{22},\quad\
U(e_1)=P_-b_+b_-\otimes f^{11}+P_-b_-b_+\otimes f^{22},\cr
U(e_2^{11})&=P_+b_-b_+\otimes f^{11}+P_+b_+b_-\otimes f^{22},\quad
U(e_2^{12})=P_+b_+\otimes f^{21}-P_+b_-\otimes f^{12},\cr
U(e_2^{21})&=P_-b_-\otimes f^{12}-P_-b_+\otimes f^{21},\quad\qquad
U(e_2^{22})=P_-b_-b_+\otimes f^{11}+P_-b_+b_-\otimes f^{22},\cr}\eqno(3.8)$$
where $f^{ij},\, i,j=1,2,$ are the matrix units on the tensor factor
${\bf C}^2$ of $\HS_\pi$ corresponding to the basis $\{f_1,f_2\}$. This
identification is correct because the vacuum vector of $\HS_\pi$ is
$H$-invariant
$$U(a)\Omega_0=\epsilon(a)\cdot\Omega_0,\qquad a\in H,\eqno(3.9a)$$
and the other vacuum vectors of the direct sum observables
$\pi(\OA)$ give rise to the defining representations of $H$:
$$U(a)\Omega_r^i=\sum_{k=1}^{n_r}\Omega_r^k\cdot D_r^{ki}(a),\qquad r=1,2,
\ i=1,\ldots,n_r.\eqno(3.9b)$$
We note that $U(e_2^{12})$ and $U(e_2^{21})$ have fermionic character. The weak
closure $\pi(\OA)^-$ of $\pi(\OA)$ contains $U(e_r), r\in\hat H$, which shows
that the fermion number has an observable meaning only on the NS sectors
$\HS_0$
and $\HS_\Fel$ because $U(e_2^{11})$ and $U(e_2^{22})$ does not belong to
$\pi(\OA)^-$.
\medskip

{\it 3.2. $H$-actions on the field algebra}\smallskip

Having a unitary realization $U$ of the symmetry algebra $H$
we would like to characterize the elements of $\BOH$ by the representations
of $H$. The maps $\lambda$ and $\rho$ defined as
$$\eqalign{H\otimes \BOH\ni a\otimes F&\mapsto \lambda_a(F):=U(a)F,\cr
           H\otimes \BOH\ni a\otimes F&\mapsto
\rho_a(F):=FU(a)\cr}\eqno(3.10a)$$
are left and right $H$-actions, respectively, on $\BOH$
therefore $\BOH$ is an $H$-bimodule. Combining $\rho$ with the antipode $S$ we
obtain a left action $\Lambda\colon H\otimes H\otimes\BOH\to \BOH$:
$$\Lambda_{a\otimes b}(F)=U(a)FU(S(b)).\eqno(3.10b)$$
One can define the left adjoint action\footnote{*}{\Fnrm But let us note that
the module algebra properties do not fulfil with this action since $H$ is not
coassociative.} $\alpha\colon H\otimes\BOH\to\BOH$
$$\alpha_a(F):=\Lambda_{\Delta(a)}(F)=U(a^{(1)})FU(S(a^{(2)}))\eqno(3.11)$$
in order to characterize the observables in an other way. If $A$ is an
observable, i.e. if it commutes with $U(H)$ then
$$U(a^{(1)})AU(S(a^{(2)}))=AU(a^{(1)})U(S(a^{(2)}))=\epsilon(a)\cdot A,
  \eqno(3.12)$$
where we used axiom 5. and the fact that $l$ is a central invertible element of
$H$, that is $U(l)$ itself is observable. Conversely, if the transformation
property $U(a^{(1)})FU(Sa^{(2)})=\epsilon(a)\cdot F,\, F\in\BOH$ holds then
$$\eqalign{U(a)[U(\varphi_1^*)FU(S(\varphi_2^*)r\varphi_3^*l)]&=
U(a^{(1)})[U(\varphi_1^*)FU(S(\varphi_2^*)S(a^{(21)})ra^{(22)}\varphi_3^*l)]\cr
&=U(\varphi_1^*)[U(a^{(11)})FU(S(a^{(12)})]U(S(\varphi_2^*)r\varphi_3^*l)
  U(a^{(22)})\cr
&=[U(\varphi_1^*)FU(S(\varphi_2^*)r\varphi_3^*l)]U(a).\cr}\eqno(3.13)$$
{}From the pentagon identity one proves that
$$\UN^{(1)}\otimes S(\UN^{(2)})=\varphi_1^*\varphi_1^{(1)}\otimes
  S(\varphi_1^{(2)})S(\varphi_2^*)r\varphi_3^*\varphi_2lS(\varphi_3),
  \eqno(3.14)$$
therefore sandwiching $F$ with both sides of (3.14) and using (2.5.a) one
obtains $F=U(\varphi_1^*)FU(S(\varphi_2^*)r\varphi_3^*l)$. Thus the commutant
of $U(H)$ is
$$U(H)'=\alpha_{e_0}(\BOH)=\pi(\OA)^-.$$
One can decompose $\BOH$ with respect to $\alpha$:
$$\BOH=\left(\bigoplus_{r\in\hat H}\CB_r\right)
  \bigoplus {\cal N},\qquad \CB_r=\alpha_{e_r}(\BOH),\ r\in\hat H,
  \quad {\cal N}={\rm Ker}\, \alpha_\UN.\eqno(3.15)$$
Therefore the left adjoint action is not enough to characterize the elements of
$\BOH$ in general since it may contain a nontrivial kernel if the coproduct is
not unit preserving.
Thus let $p_i\in H\otimes H, i=1,\ldots, n_\emptyset$ be the one dimensional
orthogonal projections in the decomposition
$$\UN\otimes\UN-\Delta(\UN)=\sum_{i=1}^{n_\emptyset} p_i=p_1+p_2\equiv
  e_2^{11}\otimes e_2^{22}+e_2^{22}\otimes e_2^{11},\qquad p_i p_j=
  \delta_{ij}\cdot p_j,\quad p_i^*=p_i,\eqno(3.16)$$
then ${\cal N}$ can be decomposed as
$${\cal N}=\bigoplus_{i=1}^{2}{\cal N}_i,\qquad
  {\cal N}_i=\Lambda_{p_i}({\cal N}).\eqno(3.17)$$
Writing a general element $\hat F\in\pi(\FA)$ in the form
$$\hat F=\left(\matrix{\hat A&\hat B\cr \hat C&\hat D\cr}\right),\eqno(3.18a)$$
where $\hat A,\ldots,\hat D$ are polynomials in NS generators with the ordering
that the four possible $b_+b_-,b_-b_+,b_+,b_-$ terms stand at the end of each
monom:
$$\hat A=A^\pm b_+b_- +A^\mp b_-b_+ +A^+ b_+ +A^- b_-,\eqno(3.18b)$$
i.e. $A^\pm,\ldots,  D^+$ do not contain already the generators $b_+,b_-$,
one checks that $\CB_0$ is just the observable algebra $\pi(\OA)^-$:
$$\alpha_{e_0}(\hat F)=
  \left(\matrix{  A_0^\pm b_+b_-&  B_1^+b_+\cr
                C_1^-b_-&  D_0^\mp b_-b_+\cr}\right)+\Fel\cdot
  \left(\matrix{(A_0^\mp+ D_0^\pm)b_-b_+&(B_1^-+  C_1^+)b_-\cr
              (B_1^-+  C_1^+)b_+&(A_0^\mp +D_0^\pm)b_+b_-\cr}\right).
  \eqno(3.19)$$
The two terms on the right hand side correspond to NS and R observables,
respectively, while the subscripts $0$ and $1$ denote even and odd parts of the
polynomials: $A^+=A^+_0+A_1^+,\ldots$, etc. The linear subspaces
$\CB_1,\CB_2,{\cal N}$ of $\BOH$ look as follows:
$$\eqalignno{\alpha_{e_1}(\hat F)&=
  \left(\matrix{  A_1^\pm b_+b_-&  B_0^+b_+\cr C_0^-b_-&  D_1^\mp b_-b_+\cr}
  \right)+\Fel\cdot
  \left(\matrix{(A_0^\mp- D_0^\pm)b_-b_+&(  B_1^- -  C_1^+)b_-\cr
  (  C_1^+-  B_1^-)b_+&(  D_0^\pm -  A_0^\mp)b_+b_-
  \cr}\right),\cr\cr
 \alpha_{e_2}(\hat F)&=
  \left(\matrix{  A^+b_+ +A^-b_-&  B^\pm b_+b_- +B^\mp b_-b_+\cr
  C^\pm b_+b_- + C^\mp b_-b_+ &  D^+b_+ +D^-b_-\cr}\right),&(3.20)\cr\cr
 \Lambda_{p_1+p_2}(\hat F)&=
  \left(\matrix{  A_1^\mp b_-b_+ &B_0^-b_-\cr
                        C_0^+b_+ &D_1^\pm b_+b_-\cr}\right).\cr}$$
{}From the subspaces $\CB_r^k=\alpha_{e_r^{kk}}(\BOH)$ one can construct the
fields
$$\CB_r^k\ni F_r^k\mapsto \tilde F_r^k\equiv U(\varphi_1^*)F_r^k
  U(S(\varphi_2^*)r\varphi_3^*l)\in\tilde\CB_r^k\eqno(3.21)$$
obeying $H$-covariant intertwining properties
$$U(a)\tilde F_r^k=\sum_{k'=1}^{n_r}\tilde F_r^{k'}D_r^{k'k}(a^{(1)})
  U(a^{(2)})\equiv\tilde F_r^{k'}U(\nu_r^{k'k}(a)),\qquad a\in H,\eqno(3.22)$$
that is they induce amplimorphisms of $\nu_r(a)=D_r(a^{(1)})\otimes a^{(2)}$
type of the symmetry algebra $H$. The elements of $\tilde\CB_r^k,\, k=1,\ldots,
n_r, r\not= 0$ are the `true' charged fields because
they map $H$-covariant states into $H$-covariant ones:
$$U(a)\cdot\tilde F_r^k\Psi_s^j=\sum_{k',j'} \tilde F_r^{k'}\Psi_s^{j'}\cdot
  (D_r\times D_s)^{k'j',kj}(a),\qquad \Psi_s^j\in U(e_s^{jj})\HS_\pi.
  \eqno(3.23)$$
The map $F\mapsto\tilde F$ in (3.21) is the identity on the observables due
to the square identity 8.b, while the fields in ${\cal N}$ are mapped into
zero due to (3.14).

\medskip
{\it 3.3. $H$-covariant multiplet fields}\smallskip

{}From the linear subspaces $\tilde\CB_r^k,\, k=1,\ldots, n_r,\, r\in\hat H$
one can consruct $H$-covariant multiplet matrix fields
$\tilde F_r=\{\tilde F_r^{\kappa k}\FU \kappa, k=1,\ldots,n_r\}\in
M_{n_r}(\BOH)$ obeying slightly modified weak F-algebra (generalized Cuntz
algebra [10]) properties [11]:
$$U(a)\tilde F_r^{\kappa k}
  =\sum_{k'=1}^{n_r}\tilde F_r^{\kappa k'}U(\nu_r^{k'k}(a)),
  \qquad \kappa=1,\ldots,n_r,\ a\in H,\eqno(3.24a)$$
$$[\tilde F_r^*\tilde F_r]^{kk'}\equiv\sum_{\kappa=1}^{n_r}
  \tilde F_r^{\kappa k*}  \tilde F_r^{\kappa k'}=
  U(\nu_r^{kk'}(\UN)),\qquad k,k'=1,\ldots n_r,\eqno(3.24b)$$
where $\nu_r$ is the $H$-amplimorphism
$$\nu_r(a)=D_r(a^{(1)})\otimes a^{(2)}.\eqno(3.25)$$
Clearly, $U(\nu_r(\UN))\in M_{n_r}(\BOH)$ is a projection, that is
$\tilde F_r\in M_{n_r}(\BOH)$ is a partial isometry. The matrix
elements $U(\nu_r^{kk'}(\UN))$ commute with the observables. In our case
$\nu_r^{kk'}(\UN)=\delta_{kk'}\cdot E_r^k$, where the projections $E_r^k\in H$
differ from the unit $\UN$ of $H$ only if the amplimorphism $\nu_r$ of $H$ is
unit non-preserving:
$$E_0=E_1=\UN,\qquad E_2^1=e_0+e_1+e_2^{11},\qquad E_2^2=e_0+e_1+e_2^{22}.
  \eqno(3.26)$$
Due to (3.24) $\tilde F_r^{\kappa k}$ obeys the following intertwining property
with an observable $A\in\pi(\OA)$ (summation is supressed):
$$\eqalign{
   \tilde F_r^{\kappa k}A&=U(\UN)\tilde F_r^{\kappa k}A
  =\tilde F_r^{\kappa k'}U(\nu_r^{k'k}(\UN))A
  =\tilde F_r^{\kappa k'}AU(\nu_r^{k'k}(\UN))\cr
 &=\tilde F_r^{\kappa k'}A\tilde F_r^{\kappa' k'*}
    \tilde F_r^{\kappa' k}
  =\left[\tilde F_r^{\kappa k'}A
    \tilde F_r^{\kappa' k'*}\right]\cdot\tilde F_r^{\kappa' k}\equiv
   \rho_r^{\kappa\kappa'}(A)\tilde F_r^{\kappa' k}.\cr}\eqno(3.27)$$
$\rho$ is an amplimorphism of the observables $\pi(\OA)$ into
$M_{n_r}(\pi(\OA)^-)$ since $\rho_r^{\kappa\kappa'}(A)$ is observable
$$\eqalign{U(a)\rho_r^{\kappa\kappa'}(A)
  &=U(a)\tilde F_r^{\kappa k}A\tilde F_r^{\kappa' k*}=
    \tilde F_r^{\kappa k'}AU(\nu_r^{k'k}(a))
    \tilde F_r^{\kappa' k*}\cr
  &=\tilde F_r^{\kappa k'}A[\tilde F_r^{\kappa' k}
    U(\nu_r^{kk'}(a^*))]^*
   =\tilde F_r^{\kappa k'}A[U(a^*)\tilde F_r^{\kappa' k'}]^*\cr
  &=\rho_r^{\kappa\kappa'}(A)U(a),\cr}\eqno(3.28a)$$
moreover
$$\rho_r^{\kappa\kappa'}(A)\rho_r^{\kappa'\kappa''}(B)=
 \rho_r^{\kappa\kappa''}(AB),\qquad
 \rho_r^{\kappa\kappa'*}(A)=\rho_r^{\kappa'\kappa}(A^*)\eqno(3.28b)$$
fulfil as well. An explicit choice of the covariant multiplet matrices is:
$$\tilde F_0=\left(\matrix{\UN&0\cr 0&\UN\cr}\right),\qquad\qquad
  \tilde F_1=\left(\matrix{
  (P_--P_+)b_-b_+&b_+\cr b_-&(P_+-P_-)b_+b_-\cr}\right),\eqno(3.29a)$$
$$\tilde F_2=\left(\matrix{
    xP_-b_+ -P_+b_-&xP_+b_+b_-&xP_+b_+ +P_-b_-&xP_-b_+b_-\cr
    xP_+b_-b_+ +P_+b_+b_-&-xP_-b_-&xP_-b_-b_+ +P_-b_+b_-&-xP_+b_-\cr
    0& P_+b_-b_+&0&P_-b_-b_+\cr
    0&P_+b_+&0&-P_-b_+\cr}\right),\eqno(3.29b)$$
where $x=1/\sqrt{2}$ and the two by two blocks in $\tilde F_2$ correspond to
the matrix elements $\tilde F_2^{\kappa k},\, \kappa, k=1,2$. Writing a general
NS and R element in $\pi(\OA)$ in the form (see (3.19))
$$O_\NS=\left(\matrix{A_0b_+b_-&  B_1b_+\cr
                C_1b_-&  D_0 b_-b_+\cr}\right),\qquad
  O_\R=\left(\matrix{E_0b_-b_+&E_1b_-\cr
              E_1b_+&E_0b_+b_-\cr}\right),\eqno(3.30)$$
the induced amplimorphisms $\rho_1$ and $\rho_2$ of the observables read as:
$$\rho_1(O_\NS)=\left(\matrix{D_0b_+b_-&  -C_1b_+\cr
                -B_1b_-&  A_0 b_-b_+\cr}\right),\qquad
  \rho_1(O_\R)=\left(\matrix{E_0b_-b_+&-E_1b_-\cr
              -E_1b_+&E_0b_+b_-\cr}\right),\eqno(3.31a)$$
$$\eqalign{\rho_2(O_\NS)&=\left(\matrix{
   A_0b_-b_+&0&0&B_1b_-\cr
   0&A_0b_+b_-&B_1b_+&0\cr
   0&C_1b_-&D_0b_-b_+&0\cr
   C_1b_+&0&0&D_0b_+b_-&0\cr}\right),\cr\cr
  \rho_2(O_\R)&=\left(\matrix{
   E_0b_+b_-&E_1b_+&0&0\cr
   E_1b_-&E_0b_-b_+&0&0\cr
   0&0&0&0\cr
   0&0&0&0\cr}\right).\cr}\eqno(3.31b)$$
Due to the choice of the multiplet matrices the images are in $\pi(\OA)$, that
is they do not contain the elements $U(e_0),U(e_1)$. Therefore (3.31) defines
amplimorphisms of the observable algebra $\OA$. The amplimorphism $\rho_2\colon
\OA\to M_2(\OA)$ is not unit preserving, but of course a left inverse exists.
The representations $\tilde \pi_r$ of $\OA$ on the (finite multiple) of the
vacuum Hilbert space $\HS_0\otimes\Co^{n_r}$, which are equivalent to the
representations $\pi_r(\OA)\equiv\pi_{\vert \HS_r}(\OA)$ are given as
$$\eqalign{\tilde \pi_r(A)(\Psi\otimes z_\kappa)&=
  \left( (\pi_0\otimes id_{n_r})(\sum_{\kappa',\kappa''=1}^{n_r}
 \rho_2^{\kappa'\kappa''}
 (A)\otimes z^{\kappa'\kappa''})\right)(\Psi\otimes z_\kappa)\cr
 &=\sum_{\kappa'=1}^{n_r}\pi_0(\rho_2^{\kappa'\kappa}(A))\Psi\otimes
z_{\kappa'}
 \cr},\eqno(3.32)$$
where $A\in\OA,\Psi\in\HS_0$, $\{ z_\kappa\}$ is an orthonormal basis in
$\Co^{n_r}$ and $z^{\kappa\kappa'}$ are the corresponding matrix units.

\bigskip
{\bf 4. Discussion and outlook}\medskip

Apart from the previously discussed chiral Ising model there is a less trivial
case [19] when a rational Hopf algebra may describe the superselection
symmetry of a chiral field theory: a rational Hopf algebra
$H=M_1\oplus M_2\oplus M_2\oplus M_1$ can reproduce the fusion rules, the
conformal weights, the quantum dimensions and the modular group representation
of the level 3 integrable representations of the $\hat A_1$ Kac--Moody algebra.
It is an example for a RHA that provides us the third smallest index
$(3+\sqrt{5})/2$ in the Jones classification [5]. Of course, there are lots of
examples for RHAs with coassociative and unit preserving coproduct: every
double ${\cal D}(G)$ of a finite group $G$ is a RHA. They describe the global
symmetries of $G$-orbifold models [20] and $G$-spin chains [11].

Thus we conjecture that the global symmetry algebras of unitary chiral
rational conformal field theories are provided by rational Hopf algebras.
In that case the classification of rational Hopf algebras leads to a partial
classification of unitary RCFTs, namely, the possible fusion rules, the
conformal weights ({\it mod} 1), the Virasoro central charge ({\it mod} 8) and
the modular group representations can be classified.

Dropping axiom 9. one can study degenerate RHAs having degenerate monodromy
matrices. They can correspond to two dimensional field theories with finite
number of superselection sectors, where not only the vacuum sector obeys
permutation group statistics [3].

Finally, we would like to mention that the low temperature behaviour of three
dimensional quantum impurity problems and the multi-channel Kondo effect can be
described in terms of chiral conformal field theories [21]. Therefore
rational Hopf algebras may emerge as symmetry algebras of an impurity atom
coupled to a three dimensional electron gas.

\bigskip
{\it Acknowledgement.} I would like to thank B. Nachtergaele and
Professor A.S. Wightman for the valuable discussions and to J. Fuchs,
A. Ganchev and K. Szlach\'anyi for the valuable comments.

\vfill\eject

{\bf References}\bigskip

\item{[1]} S. Doplicher, R. Haag and J.E. Roberts, Commun.
Math. Phys. {\bf 13} (1969) 1; ibid {\bf 15} (1969) 173; ibid
{\bf 23} (1971) 199; ibid {\bf 35} (1974) 49\smallskip

\item{[2]} S. Doplicher and J.E. Roberts, Bull. Am. Math.
Soc. {\bf 11} (1984) 333; Ann. Math. {\bf 130} (1989) 75; Invent. Math.
{\bf 98} (1989) 157; Commun. Math. Phys. {\bf 131} (1990) 51\smallskip

\item{[3]} K.-H. Rehren, {\sl Braid group statistics and their
superselection rules}, in: Algebraic theory of superselection sectors,
ed. D. Kastler, (World Scientific 1990) p. 333 \smallskip

\item{[4]} R. Haag, {\sl Local quantum physics} (Springer 1992)\smallskip

\item{[5]} V. Jones, Invent. Math. {\bf 72} (1982) 1\smallskip

\item{[6]} M. Pismer and S. Popa, Ann. Sci. Ecole Norm. Sup. {\bf 19}
(1986) 57;

\item{} Y. Watatani, Memoirs of the AMS, No. 424 (1990)\smallskip

\item{[7]} K. Fredenhagen, {\sl Generalizations of the theory of
superselection sectors},
in: Algebraic theory of superselection sectors, ed. D. Kastler,
(World Scientific 1990) p. 379\smallskip

\item{[8]} G. Mack and V. Schomerus, Commun. Math. Phys. {\bf 134}
(1990) 139\smallskip

\item{[9]} R. Brunetti, D. Guido and R. Longo, {\sl Modular structure and
duality in conformal quantum field theory}, Universita di Roma preprint (1992);

\item{} F. Gabbiani and J. Fr\"ohlich, {\sl Operator algebras and conformal
field theory}, ETH-preprint TH/92-30 (1992)\smallskip

\item{[10]} J. Cuntz, Commun. Math. Phys. {\bf 57} (1977) 173\smallskip

\item{[11]} K. Szlach\'anyi and P. Vecserny\'es, {\it Quantum symmetry and
braid group statistics in $G$-spin models}, KFKI-Budapest preprint 1992-08/A
(to appear in Commun. Math. Phys.)\smallskip

\item{[12]} K. Szlach\'anyi and P. Vecserny\'es, {\sl Rational Hopf algebras}
(to be published)\smallskip

\item{[13]} V.G. Drinfeld, St. Petersburg Math. J. {\bf 1} (1990) 1419
\smallskip

\item{[14]} G. Mack and V. Schomerus, Phys. Lett. {\bf B267} (1991) 207;

\item{} G. Mack and V. Schomerus, Nucl. Phys. {\bf B370} (1992) 185\smallskip

\item{[15]} J. Fuchs, {\sl Affine Lie algebras and quantum groups} (Cambridge
University Press 1992) \smallskip

\item{[16]} V. Schomerus, {\sl Quantum symmetry in quantum theory},
DESY-preprint 93-018 (1993)\smallskip

\item{[17]} M.E. Sweedler, Hopf algebras (W.A. Benjamin Inc.,
New York 1969)\smallskip

\item{[18]} H. Araki, Contemporary Mathematics {\bf 62} (1987) 23\smallskip

\item{[19]} J. Fuchs, A. Ganchev and P. Vecserny\'es, (in preparation)
\smallskip

\item{[20]} R. Dijkgraaf, C. Vafa, E. Verlinde and H. Verlinde,
Commun. Math. Phys. {\bf 123} (1989) 485;

\item{} P. B\'antay, Phys. Lett. {\bf B245} (1990) 475\smallskip

\item{[21]} A.W.W. Ludwig and I. Affleck, Phys. Rev. Lett. {\bf 67} (1991) 161,
3160 \smallskip

\vfill\eject
\end